\documentclass[useAMS,usenatbib]{mn2e}
\usepackage{graphicx}

\title[An ultradeep submillimetre map]{An ultradeep submillimetre map: beneath the SCUBA confusion limit with lensing and robust source extraction}
\author[K. K. Knudsen, V. E. Barnard et al.]
   {K.~K.~Knudsen,$^1$
    V.~E.~Barnard,$^3$ P.~P.~van~der~Werf,$^1$
    P.~Vielva,$^4$ J.-P.~Kneib,$^{5,6,7}$ \newauthor A.~W.~Blain,$^6$ R.~B.~Barreiro,$^4$ 
    R.~J.~Ivison,$^{8,9}$ I.~Smail,$^{10}$ and J.~A.~Peacock$^9$ \\ 
    $^1$Leiden Observatory, P.O.Box 9513, NL-2300 RA Leiden, The Netherlands \\
    $^2$Max-Planck-Institut f\"ur Astronomie, K\"onigstuhl 17, D-69117 Heidelberg, Germany \\
    $^3$Joint Astronomy Centre, 660 North A`oh\={o}k\={u} Place, University Park, Hilo, HI 96720, USA \\
    $^4$Instituto de F{\'\i}sica de Cantabria, Av. de Los Castros s/n, 39005, Santander, Spain\\
    $^5$Observatoire Midi-Pyr\'en\'ees, UMR5572, 14 Avenue Edouard Belin, F-31000 Toulouse, France \\
    $^6$California Institute of Technology, MS 105-24, Pasadena, CA 91125, USA\\
    $^7$Laboratoire d'Astrophysique de Marseille, OAMP, Traverse du Siphon - B.P.8, 13376 Marseille Cedex 12, France\\
    $^8$Astronomy Technology Centre, Royal Observatory, Blackford Hill, Edinburgh, EH9 3HJ, UK\\
    $^9$Institute for Astronomy, Royal Observatory, Blackford Hill, Edinburgh EH9 3HJ, UK \\
    $^{10}$Institute for Computational Cosmology, Department of Physics, Durham University, South Road, Durham DH1 3LE, UK}
\date{}

\pagerange{\pageref{firstpage}--\pageref{lastpage}} \pubyear{2005}

\def\LaTeX{L\kern-.36em\raise.3ex\hbox{a}\kern-.15em
    T\kern-.1667em\lower.7ex\hbox{E}\kern-.125emX}

\begin{document}

\label{firstpage}

\maketitle

\begin{abstract}
Extracting sources with low signal-to-noise from maps with structured 
background is a non-trivial task which has become important in studying the faint end 
of the submillimetre number counts.  In this article we study source 
extraction from submillimetre jiggle-maps from the 
Submillimetre Common-User Bolometer Array (SCUBA) using 
the Mexican Hat Wavelet (MHW), an isotropic wavelet technique. 
As a case study we use a large (11.8\,arcmin$^2$) jiggle-map of the galaxy cluster 
\mbox{Abell\,2218}, with a 850-$\mu$m 1$\sigma$ r.m.s.\ sensitivity of 0.6-1\,mJy.    
We show via simulations that MHW is a powerful tool for reliable extraction of 
low signal-to-noise sources from SCUBA jiggle-maps and nine sources 
are detected in the A2218 850-$\mu$m image.  Three of these sources are
identified as {\rm images of a single background source with an unlensed flux of} 0.8\,mJy.  
Further, two 
single-imaged sources also have unlensed fluxes $<2$\,mJy, below the 
blank-field confusion limit.    
In this ultradeep map, the individual sources detected resolve 
nearly all of the extragalactic background light at 850\,$\mu$m, 
and the deep data allow to put 
an upper limit of 44 sources per arcmin$^2$ to 0.2\,mJy at 850\,$\mu$m. 

\end{abstract}


\section{Introduction}  

The arrival of the Submillimetre Common-User Bolometer Array (SCUBA; 
\citealt{holland}) on the James Clerk Maxwell Telescope (JCMT) 
heralded a new era in galaxy evolution studies.  
The populations of submillimetre-bright, high redshift galaxies 
\citep{smail97a,hughes98,barger98} have been studied using three different 
observational strategies:
gravitational lens surveys, where pointings are made in the regions of 
well known cluster lenses hence benefiting from the cluster magnification,  
blank field surveys, generally selected to overlap with areas already 
observed at other wavelengths,  and pointed photometry of targets 
selected in other bands. 

The blank field surveys 
(\citealt{barger}; \citealt{hughes98}; \citealt{eales}; \citealt{scott_8mJy}; 
\citealt{borys03}; \citealt{webb_3h}) have produced 
counts at fluxes brighter than $S_{850}\sim$\,2\,mJy.  This 
flux limit is caused by the source confusion, 
which for the 15$''$ angular resolution of the JCMT at 850\,$\mu$m 
appears to become a problem at about 2 mJy, effectively limiting the
depths of these surveys (\citealt{condon74}; \citealt{blainconf}; 
 \citealt{barger}; \citealt{eales}; \citealt{hogg}). 
To this flux limit 35-45\% (e.g., \citealt{borys03}) of the submillimetre 
extragalactic background light \citep{puget96,fixsen98} is resolved. 

In comparison, the cluster gravitational lens surveys 
(\citealt{smail97a,sibk02}; \citealt{cowie}; 
\citealt{chapman02}; Knudsen et al., {\it in prep.}),
probe sources fainter than the blank field confusion limit. 
As the lensing transformation is a mapping from source plane to image 
plane, it changes the number density of observed sources:
Both the flux of a given source and the area of the region surveyed 
are magnified. 
As a result the confusion limit is 
moved to a fainter flux level, thereby enabling us to reliably 
study the fainter sub-mm galaxy population, which would otherwise not be
accessible with the current telescopes and instrumentation.  
The magnification can be very large near critical lines, where
we expect to see multiple images of the background sources \citep{kneib96}. 
The typical size of the critical lines is 20-30\arcsec\ in radius for 
massive clusters and hence SCUBA's 2.3\arcmin\ diameter field-of-view is very 
well-matched to the high magnification region of typical cluster lenses.

This faint population, which can currently only be probed via lensing, 
is crucial to study as it is an important contributor to the 
extragalactic submillimetre background light 
(e.g., \citealt{blain97,blain99,borys03,KnudsenPHD}).  
While many tens of submillimetre galaxies with fluxes $>$4\,mJy have 
already been studied in great detail probing their redshift distribution, 
gas content etc. (e.g., \citealt{chapman05,greve05}), very 
little is known about the faint population as it yet has been possible 
only to identify and study a handful of objects in detail (e.g., 
\citealt{Kneib_a2218,borys04}).

In this paper we present one of the 
deepest ever SCUBA maps of a massive galaxy cluster:
Abell 2218,  a rich galaxy cluster at redshift $z=0.175$ 
\citep{abell89} and acts as a powerful gravitational lens
both distorting and magnifying the distant galaxies \citep{kneib96}.
It has already proven to be a powerful tool in studying 
high redshift, faint galaxies.  Exploiting the gravitational lensing, 
several background galaxies have been found, including the 
lensed $z=5.57$ galaxy of \citet{ellis01}.  Probably the most unusual 
galaxies have been a multiply-imaged submillimetre galaxy seen 
as three distinct images \citep{Kneib_a2218}, 
which have already been studied in great detail at other wavelengths 
\citep{kneib05,sheth04,garrett05}, and a redshift $z\sim7$ galaxy 
\citep{Kneib04b,egami05}. 

To analyse this new SCUBA map, we used a novel technique
based on the Mexican Hat Wavelet (MHW, so-called because the
mother wavelet is like a sombrero in \mbox{3-d} space) technique and routines
developed by \citet{cayon}, \citet{vielva1,vielva2}  and \citet{vielva3} for 
use with anticipated 
\emph{Planck Surveyor}\footnote{http://www.esa.int/science/planck} data.  
In the analysis of the \emph{Planck} data MHW will be used for removing
contaminating foreground 
extragalactic point sources and leave the Cosmic Microwave Background 
(CMB) signal.  However, in this paper we are interested in the point 
sources themselves as the end product.  The key advantage of the MHW 
technique is that without needing to characterise the background and 
noise in an image, Gaussian shaped beams can be picked out from a 
variety of backgrounds with accurate positions and fluxes.  This makes 
the technique ideal for submillimetre images where sources are significant 
in size with respect to the field-of-view, and the noise is often relatively 
high, variable across the field and hard to model accurately.
\citet{barnard04} successfully applied MHW as a source extraction 
tool on SCUBA scan-maps of Galactic regions and placed 
valuable upper limits for the source counts of 
submillimetre galaxies at 50 and 100\,mJy. 

This paper is Paper I in a series of two papers and is an introduction to the 
use of the MHW technique in SCUBA jiggle-maps followed by its application on the
deep Abell 2218 SCUBA map. 
The SCUBA data presented here include more observations than were 
published for the same cluster in \citet{Kneib_a2218}.  
Paper II (Knudsen et al., in preparation) 
presents the multi-wavelength identification of the underlying galaxies giving 
rise to the submillimetre emission. 
The present paper is organized as follows: 
Section 2 covers the observations and the data
reduction. In Section 3 we discuss in details the MHW technique.
The detailed analysis of the \mbox{Abell\,2218} sub-mm maps, the detected
sources and derived number counts are discussed in section 4.
Conclusions and prospects are summarised in section 5.
In a further paper (Knudsen et al., {\it in prep.})
we will describe the results of applying the MHW technique to a wider survey of 
cluster lenses (the Leiden-SCUBA Lens Survey). 

Throughout we will assume an $\Omega=0.3$, $\Lambda=0.7$
cosmology with $H_0=70$\,km\,s$^{-1}$\,Mpc$^{-1}$. 


\section{Observations and data reduction}    

\subsection{Observations}
\label{subsect2:obs}

\begin{figure}
\center{\includegraphics[width=7cm]{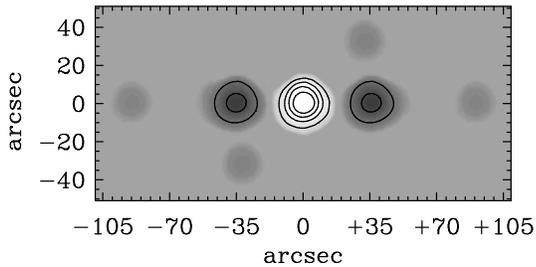}}
\caption[]{
The 850\,$\mu$m beam map.  The contours represent the levels 
$-$0.3 to 0.9 in steps of 0.2. 
The data for A2218 were obtained with different chopping configurations. 
The beam map was constructed by weighting the different chop configurations 
with observing time.  
Most of the data were obtained with a chop throw of 35\arcsec.  
The configurations with chop throws of 45\arcsec\ and 90\arcsec\ were used for a 
short time in the observations, which is reflected 
by the few per cent in the beam map.  
The negative dips at 35\arcsec\ have a little tail 
away from the central peak, which is due to the 45\arcsec\ chop throw.  
\label{fig2:beam}}
\end{figure}

We obtained observations of the cluster of galaxies A2218 
at 850\,$\mu$m  and 450\,$\mu$m with SCUBA 
during March 1998, August 2000, January 2001 and January 2002. 
SCUBA is a submillimetre mapping instrument operating at 850\,$\mu$m and 
450\,$\mu$m simultaneously (\citealt{holland}). 
SCUBA  consists of two arrays of bolometers having 
37 elements at 850\,$\mu$m and 91 elements at 450\,$\mu$m.
The field--of--view on the sky, which is approximately the same for 
both arrays, is roughly circular with a diameter of 2.3\arcmin. 
The observations were carried out in jiggle mode, where the secondary mirror follows a 64-point jiggle 
pattern in order to fully sample the beam at both operating wavelengths. 
In order to cover a larger sky area four pointings were made, though with
a big overlap region. 
As the region with large gravitational lensing magnification is extended 
due to the double-peaked mass distribution of A2218, four pointings were 
obtained.  The offsets between the central position of each pointing were about 0.5\,arcmin.
The primary sky subtraction is achieved by chopping the secondary mirror. 
The chop configuration used for most of the time was a chop throw of 
35\arcsec\ with a position angle fixed in right ascension (RA), 
though larger chop throws and other position angles were also used 
for a limited part of the observations.  The resulting beam, weighted by time spent using the different chop throws, is shown in 
Figure \ref{fig2:beam}. 
As a result of chopping to either side 
the beam pattern has a central positive peak 
with a negative dip on each side in the chop direction with amplitudes of half the peak flux. 

During the observations the pointing was checked every hour by observing 
nearby bright blazars with known positions.  
The noise level of the bolometer arrays was 
checked at least twice during an observing shift, and the atmospheric opacities at 850\,$\mu$m and 450\,$\mu$m, 
$\tau_{850\mu{\rm m}}$and $\tau_{450\mu{\rm m}}$, were determined by a JCMT skydip observation
every two--three hours and supplemented with the $\tau_{225{\rm GHz}}$ data 
from the neighboring Caltech Submillimeter Observatory (CSO). The zenith opacity was 0.15\,$<\tau_{850\mu{\rm m}}<$\,0.40 
(corresponding to 0.037\,$<\tau_{225{\rm GHz}}<$\,0.1), although data taken when $\tau_{850\mu{\rm m}}>$\,0.32 were not included in the 450-$\mu{\rm{m}}$ map since it is much more sensitive to poor weather.  
Calibrators with accurately known flux densities 
were observed every two to three hours.  If available, primary 
calibrators, i.e.\ planets, preferably compact Uranus, were observed 
at least once during an observing shift. The uncertainty in the flux 
calibration is approximately 10\,\% at 850\,$\mu$m and approximately 
30\,\% at 450\,$\mu$m. 

The on-sky exposure time was 39.7 hours at 850\,$\mu$m and 33.4 hours at 
450\,$\mu$m without overhead, i.e.\ without time needed for the chopping, 
jiggle, etc.  The total area surveyed is 11.8\,arcmin$^2$ at 850\,$\mu$m 
and 9.1\,arcmin$^2$ at 450\,$\mu$m.

\begin{figure*}
\hskip-4mm\includegraphics[width=13.4cm]{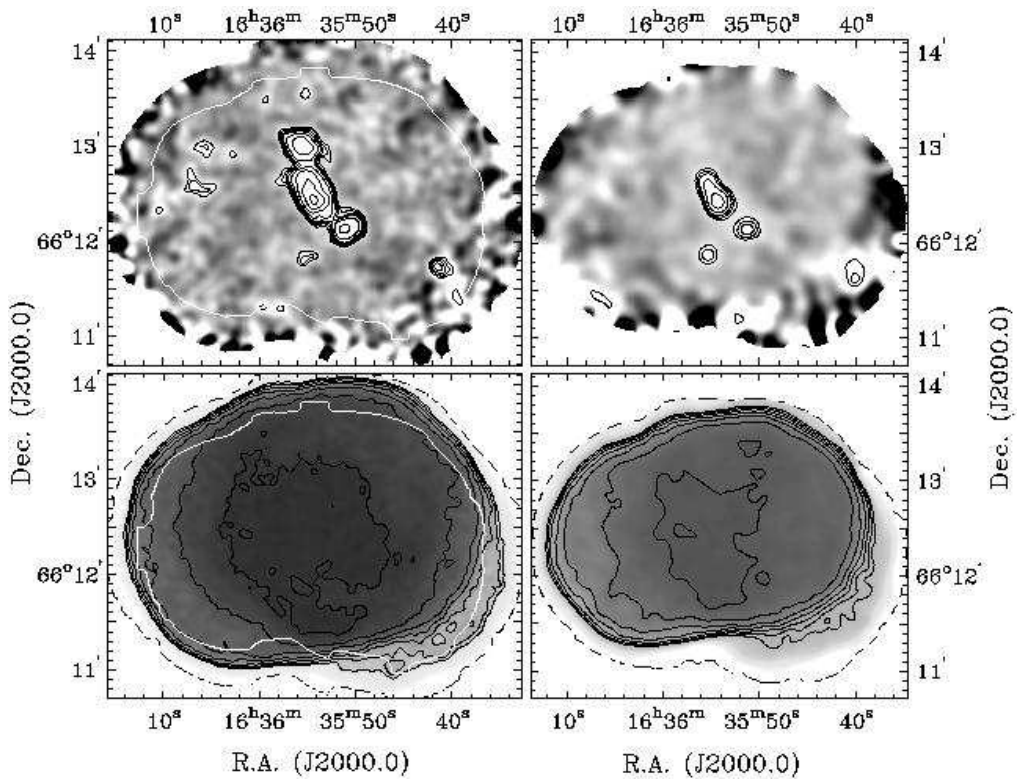}
\caption[Mosaic of the 850 and 450 data]{
In this mosaic the SCUBA flux maps (top) of A2218 and the corresponding 
noise maps are shown (bottom).  
The left side represent the 850\,$\mu$m maps, and the right side 
the 450\,$\mu$m maps.  
{\rm Top-left}:\ The 850\,$\mu$m flux map overlaid with 
850\,$\mu$m signal-to-noise contours of $S/N =\,$3,4,5,6,8,12,18,24.  
The map has been cleaned and sources have been restored using the central 
positive peak of the beam (see Sect.\ \ref{subsect2:clean}).  
The map has been smoothed with a two-dimensional Gaussian with width 5\arcsec. 
The white line indicate the area being used in the further analysis. 
{\rm Top-right}:\ The 450\,$\mu$m flux map overlaid with the 
450\,$\mu$m signal-to-noise contours of $S/N =\,$3,4,5,7,10. 
The map has been smoothed with a two-dimensional Gaussian with width 10\arcsec. 
{\rm Bottom-left}:\ The 850\,$\mu$m noise map.  
This map represents the position dependent noise after smoothing with 
5\arcsec. The contours have been overlaid to enhance the contrast and are 
0.8 to 3.6\,mJy/beam in steps of 0.4\,mJy/beam. 
Darker shades of grey indicate a lower noise. 
The dashed line indicates the area of the map.  
The edges are clearly less sensitive and the whole edge region of the map is 
not included in the analysis.  
The white line indicate the area being used in the further analysis. 
{\rm Bottom-right}:\ The 450\,$\mu$m noise map.  
This is the noise map for the data smoothed with 10\arcsec. 
The contours have been overlaid to enhance the contrast and represent 
4 to 16\,mJy/beam in steps of 2\,mJy/beam. 
The dashed line indicates the area of the map.  
Comparing left to right, the slightly smaller field of view at 
450\,$\mu$m is evident.  
\label{fig2:850450mos}}
\end{figure*}

\subsection{Reduction}
\label{subsect2:red}

The data were reduced using the standard {\sc surf} package 
(Jenness \& Lightfoot 1998). 
The reduction procedure is described in detail in the catalogue 
paper for the Leiden-SCUBA Lens Survey \citep{KnudsenPHD}. 
The maps have been regridded with a pixel scale of $1''$. 
The full width at half maximum (FWHM) of the beam is 14.3\arcsec\ in 
the final 850\,$\mu$m and 7.5\arcsec\ in the 450\,$\mu$m.  

Only the beam scale and above is real information reflected.  Thus 
to suppress artificial pixel noise, the maps are smoothed to reduce 
the spatial high frequency noise.  
The 850\,$\mu$m and the 450\,$\mu$m maps were smoothed using a 2-D 
Gaussian function with a FWHM of 5\arcsec\,, yielding the beams of 
15.1\arcsec\ and 9\arcsec\ respectively.  
Additionally, to match the beamsize of the 850\,$\mu$m, 
the 450\,$\mu$m map is smoothed with a 2-D Gaussian 
with FWHM of 10\arcsec\,, which results in a beam of 12.5\arcsec\,, 
comparable to the beam size at 850\,$\mu$m. 
The resulting maps are shown in Figure \ref{fig2:850450mos}. 
The sensitivity in the final 850\,$\mu$m map is 0.65\,mJy/beam in the 
deepest part of the map and 1.1\,mJy/beam averaged across the map.  
This is of equivalent depth to the 850-maps published by \citet{cowie}. 
The sensitivity in the final 450\,$\mu$m map is 4.3\,mJy/beam 
in the deepest part and 14\,mJy/beam averaged across the map, 
as determined from the noise simulations described below. 

In this paper we concentrate on the 850\,$\mu$m map.  
In general the analysis of 450\,$\mu$m maps is more difficult because 
the 450\,$\mu$m beam pattern is very sensitive to temperature 
variations on the primary mirror of the JCMT.  The beam pattern cannot easily 
be described by a 2D Gaussian function.  Furthermore, both the atmospheric 
opacity and the uncertainty in the flux calibration are large at 
450\,$\mu$m.  This makes observations of faint objects difficult. 
However, given the good quality of the 450\,$\mu$m data obtained for 
A2218 we do include the 450\,$\mu$m flux for the 850\,$\mu$m sources 
(Table~\ref{tab2:results}).

\subsection{Construction of noise maps}
\label{subsect2:noisemaps}
 
When a data set is combined and regridded into a map, some areas are
sampled less than the rest of the map.  In the A2218 images, these include the areas along the edges,  
the areas where the bolometers are either more noisy or where data has been 
excluded, and areas with less integration time outside the overlap 
between different pointings.  It is thus imperative to assess the noise 
as function of position in the map to be able to make a reliable 
analysis of the observed flux map. 

When data files for the SCUBA jiggle maps have been fully reduced, 
i.e.~all correlated noise has been removed, the noise in each individual 
bolometer can be assumed to be independent of other bolometers.  
The sources we are observing are 
very faint and thus large integration times are necessary in order to 
reach the needed noise levels for detections. However, the observations 
are carried out on short time scales on the order of an hour, which are the 
standard lengths of our observations, thus the data stream from the 
bolometers are dominated by noise rather than signal.  

In order to determine the position--dependent noise a Monte Carlo 
simulation was used to construct `empty' (i.e.\ no source signal present) 
maps representing the noise component in the real data files.
For each original data file, the statistical properties of each 
bolometer's data set were calculated.  Then a simulated data set for each 
bolometer was created using a Gaussian random number generator with the 
same properties.  This produced an empty map for each original data file, 
and the entire set of these was reduced and concatenated using {\sc surf} 
in the same manner as the original data files.  The resultant single map 
is thus one realization of the empty map simulation procedure.  This 
process was then repeated about 500 times.  
Noise maps were constructed from the simulated empty maps by taking all 
the empty maps and calculating the standard deviation at each pixel. 
The noise maps at both 850\,$\mu$m and 450\,$\mu$m are
shown in Fig.~\ref{fig2:850450mos}, 
where it can be seen how the noise is not uniform across the field 
and how the different effects described above appear.   

This algorithm is similar to looking at the scatter of the data that goes 
into each pixel.  It does not take into account residual 1/f 
noise and other systematic errors.  Comparing the two methods suggests 
the difference is only at the 10\% level (C.\ Borys, private communication). 

\subsection{Deconvolution of beam map}
\label{subsect2:clean}

As described in section \ref{subsect2:obs}, the chopping of the 
secondary mirror on either side of the central pointing creates a beam pattern 
with a positive peak and two negative sidelobes.  
This is undesirable because negative sidelobes can hide sources 
or bias their fluxes.  Also the negative sidelobes contain useful signal 
that can be used in source detection.  Therefore it is useful to deconvolve 
the beam pattern out.  Since we are dealing mostly with isolated point 
sources the classical CLEAN algorithm (\citealt{hogbom}) is ideally suited for 
this purpose. 
In summary, the classical CLEAN algorithm is an iterative algorithm which 
searches for the brightest point, which matches the beam pattern, 
in the final map.  
The beam pattern is scaled to some fraction of the peak flux and subtracted
from the map at that position.  
The whole process is then repeated in principle until the 
residual map has an r.m.s.\ comparable with the noise, 
though in practice until a given flux limit. 
All the information about subtracted fluxes and positions (the resultant 
``delta-functions'') can then be used for source detection and restoration 
with a different beam pattern e.g.\ the central positive peak of the 
beam pattern. 

In our cleaning algorithm, 
the (cleaned) data map was convolved during each iteration with the beam 
map and divided by the associated convolved noise map.  The former step enhances any present signal in the map whilst the latter ensures that no 
noise peaks (e.g. along the edges) would be mistaken for signal.  This image is then used to select the peaks, but the cleaning 
itself takes place on the unconvolved data map.  We found that subtracting 
10\,\% of the flux at each iteration produced acceptable convergence.  
The cleaning was continued until the residual map contained no pixels  
brighter than three times the r.m.s.\ noise level. 
Finally, the detected sources were convolved with the positive peak 
of the beam map and coadded to the uncleaned residuals to give a final 
cleaned map.
Using the central positive peak to restore 
the sources in the map preserves the information about the beam shape, which 
will be important for the source extraction.  
The cleaned 850\,$\mu$m map, where the sources have been restored 
using the central positive peak of the beam, 
is shown in Figure \ref{fig2:850450mos}.  The 450\,$\mu$m map has been 
cleaned in the same manner and is also shown in \ref{fig2:850450mos}. 


\section{Mexican Hat Wavelet Source Detection}   
\label{sect2:mhw}

Independent of the survey strategy, the source extraction algorithm 
applied to the SCUBA maps must be a robust, well-characterized method.
Indeed, the noise in SCUBA jiggle maps has a temporal variation and the 
sources we wish to detect typically have at best moderate signal-to-noise 
ratio ($S/N$).  The far-infrared emitting regions in high redshift sources 
are expected to be relatively compact and so are likely to be unresolved in 
SCUBA's 15\arcsec\ beam. 
While Mexican Hat Wavelet source detection is the scope 
of this paper, we first summarise a number of techniques have been used 
for locating point sources in previous SCUBA jiggle-maps.  

In the cluster lens surveys, extraction techniques have tended to be 
simpler, as these surveys cover smaller fields with fewer sources.  
\citet{smail97a} and \citet{sibk02} used SExtractor 
(\citealt{bertin}), which was primarily developed for use in optical 
images which have very different noise characteristics to SCUBA maps. 
In the Hawaiian deep fields, \citet{barger} simply used a signal-to-noise 
criterion for their source selection, a technique also used for the 
HDF \citep{hughes98}.

For the ongoing SCUBA Half-Degree Extragalactic Survey (SHADES) a 
matched-filter source-extraction technique has been demonstrated for a 
sub-set of data \citep{mortier05}. 

In their jiggle-maps of the Canada UK Deep SCUBA Survey (CUDSS) fields, 
\citet{eales} and \citet{webb_3h}
selected sources from a convolved signal-to-noise map, which was 
constructed by convolving their 
reduced map with a beam map derived from a calibrator,  and dividing 
by a simulated noise map (see section \ref{subsect2:noisemaps}). 
An iterative deconvolution (CLEANing) routine was then used 
for determination of the  position and flux of the detected sources. 

In the 8-mJy survey (\citealt{scott_8mJy}), sources were extracted by a 
simulta\-neous maximum-likelihood fit to the flux densities of all 
probable peak locations, selected purely on the grounds of flux.  A 
template beam map from a calibration observation was centred on every 
pixel with flux $>\, $3\,mJy in a map convolved with a Gaussian filter.  
The height of each potential source was then increased independently 
until a minimum was found in the $\chi^{2}$ statistic between this 
built-up map and the real map.  A similar approach was used in the 
extended Hubble Deep Field (HDF) and flanking fields source extraction 
(\citealt{serjeant}).  
In short, in the two methods applied to the CUDSS and to the 8-mJy 
survey data sets the images are convolved with the beam.  The methods 
differ in how the convolved images are converted to a catalogue.

\subsection{MHW Detection parameters}
\label{subsect2:detpars}

The standard mathematics and detection parameters involved in the Mexican Hat 
Wavelet (MHW) transform were given in \citet{barnard04}.  
Further details can also be found in the original 
papers describing this work (\citealt{cayon}; 
\citealt{vielva1,vielva2}).  

Point source detection is primarily controlled by setting numerical
requirements for two parameters, namely an optimal scale, $R_{opt}$, 
and an wavelet coefficient value, $w_f(R_{opt})$. 
We performed initial, simple simulations, using both real data and the
simulated noise maps from the previous section, to understand the response
of the MHW routines to SCUBA jiggle-maps, since these are quite different
in several respects to the simulated $Planck$ data sets on which the
technique was previously tested.  The procedure involved in selecting
sources for the final detection list is as follows:

\begin{enumerate}
\item Firstly the optimal scale, $R_{opt}$, is calculated by the MHW
software for each input map.  This involves iterating through small 
changes in the value of $R$ around the point source scale, $\theta$, 
until the maximum amplification is found for the map.
The final value will be dependent upon the measured impact of the
noise on varying scales (\citealt{vielva2}) ---
noise with a characteristic scale a little larger than $\theta$,
for instance, can be most strongly counteracted
by an optimal scale $R_{\rm{opt}}$ slightly smaller than $\theta$.
\item Point source candidates are then selected at positions with wavelet
coefficient values $w_f(R_{opt}) \geq 2\sigma_{w_n}(R_{opt})$, where 
$\sigma_{w_n}(R_{opt})$, where $\sigma_{w_n}(R)$ is the dispersion of 
the noise field in wavelet space.
The value of 2 was suggested by our early simulations as a value which
allowed all real sources to pass through to an initial catalogue.
\item For each candidate, the `experimental' $w_f(R)$ is compared to the
theoretical variation expected with $R$, as a further check on the source's
shape.  A value of $\chi^2$ is calculated between the expected and experimental
results and the second parametrised requirement is therefore that
$\chi^2 \leq 4$, i.e. that the region surrounding the identified peak has
the characteristics of a Gaussian point source.
\end{enumerate}

Since, as discussed in Section \ref{subsect2:noisemaps}, the noise level in 
the map varies widely, the input map to the MHW software used was a 
signal-to-noise map, created by \mbox{dividing} the final, smoothed data map by 
the final, smoothed noise map.  This normalises the input map with respect 
to the noise.  This is especially useful for this technique as the 
properties of the whole map are used to determine detection parameters, 
and so high noise values can distort these values unfairly.  

Despite this normalisation, detections close to the edge of the map, 
where the noise levels are at their highest, were considered to still 
be unreliable and such detections in a 1.5-beam region around the edge 
are not retained in the final catalogue.
As discussed in previous papers (e.g., \citealt{ivison02}) the 
number of spurious detections in SCUBA maps is larger along the edges 
than elsewhere in the maps. 

To assess the chances of obtaining false positive detections in the $S/N$ map, 
we performed the MHW source extraction upon empty $S/N$ maps, created by 
dividing the empty Monte Carlo maps by the final average noise map.  In 100 
such realisations we found only two sources with $S/N >$ 3.  We thus conclude 
that, when restricting the sources to those which both meet the standard 
detection parameter requirements outlined and which have 
$S/N >$ 3 in real space, the probabililty of spurious detections is very low.   
A detailed analysis of this is performed for the 
full Leiden-SCUBA Lens Survey (Knudsen, 2004; Knudsen et al., {\it in prep.}). 

\begin{table}
\caption{MHW and CLEAN detections}
\label{tab:mhwclean}
\begin{center}
\begin{tabular}{rr@{\hspace{1mm}}r@{\hspace{3mm}}rr@{\hspace{1mm}}r@{\hspace{3mm}}rr@{\hspace{1mm}}r@{\hspace{3mm}}r}
\hline
\hline
 & \multicolumn{6}{c}{MHW} & \multicolumn{3}{c}{CLEAN} \\
 & \multicolumn{3}{c}{uncleaned} & \multicolumn{3}{c}{cleaned} & \multicolumn{3}{c}{} \\
src & $\Delta x$ & $\Delta y$ & \multicolumn{1}{c}{$f$} & $\Delta x$ & $\Delta y$ & \multicolumn{1}{c}{$f$} & $\Delta x$ & $\Delta y$ & \multicolumn{1}{c}{$f$} \\
\hline
  1 &  78 & -40 & 10.8  &  79 & -40 & 10.4 &    78 & -41 &  9.0 \\
  2 &  20 & -17 &  9.0  &  20 & -17 &  8.7 &    18 & -17 &  9.2 \\
  3 &  -1 &   2 & 17.1  &  -0 &   1 & 16.1 &    -1 &   0 & 22.3 \\
  4 &  -6 &  15 & 11.7  &  -6 &  14 & 12.8 &    -8 &  14 & 11.3 \\
  5 &  -6 & -34 &  5.0  &  -6 & -34 &  3.1 &    -7 & -34 &  3.2 \\
  6 &  -8 &  36 & 11.0  &  -8 &  36 & 11.3 &   -11 &  36 & 13.9 \\
  7 & -54 &  30 &  4.3  & -51 &  31 &  3.3 &   ... & ... &  ... \\
  8 & -69 &  35 &  5.4  & -69 &  35 &  5.2 &   -68 &  32 &  3.8 \\
  9 & -75 &  10 &  4.6  & -74 &  10 &  4.8 &   -75 &  10 &  4.4 \\ [0.1ex]
 10 &  45 &  13 &  4.4  & ... & ... & ...  &   ... & ... &  ... \\
 11 &  59 & -17 &  4.0  & ... & ... & ...  &   ... & ... &  ... \\
 12 & 113 &  -1 &  9.3  & ... & ... & ...  &   ... & ... &  ... \\
 13 & ... & ... & ....  & ... & ... & ...  &   -30 &  68 &  2.0 \\
 14 & ... & ... & ....  & ... & ... & ...  &   -10 &  69 &  2.3 \\
 15 & ... & ... & ....  & ... & ... & ...  &     2 &  29 &  2.4 \\
 16 & ... & ... & ....  & ... & ... & ...  &    30 &  50 &  2.6 \\
\hline
\end{tabular}
\end{center}
Note: \  $\Delta x$ and $\Delta y$ give the distance 
in pixels from the map centre at $\alpha,\delta$ = 
16$^{\rm h}$35$^{\rm m}$54.22$^{\rm s}$,+66$^\circ$12\arcmin24\arcsec, 
and the pixel scale is 1 arcsec/pixel.  The flux, $f$, is in units mJy. 
\end{table}

\subsection{Comparison of MHW and CLEAN source detection} 
 
We compare the MHW source detection done on the un-cleaned map and 
the cleaned map.  x,y pixel positions relative to the centre of the map 
and fluxes for detections are listed in Table~\ref{tab:mhwclean}.
The first nine sources are numbered according to their number
in the final catalogue as given in Table~\ref{tab2:results}.  
The positions agree very well 
for all nine sources, though for source 7 there has been a four 
pixel shift.  Source 7 is close to the negative sidelopes of 
the bright central sources.  In the cleaned map its wavelets parameters
are much improved and its profile provides a better match to the theoretical 
expectation as estimated by the MHW algorithm.  Source 7 is the 
reason why we CLEANed the SCUBA map for this project, i.e., 
to confirm whether the source was an artifact due to the 
negative sidelopes of the beam.  For the other sources the wavelets 
parameters were improved only by a few percent when using the 
cleaned map.   In the uncleaned map three additional sources were 
detected.  Source 12 falls within the edge region, which has been 
trimmed off.  Sources 10 and 11 do not produce a good match to the 
theoretical wavelets expectation.  As these sources were not 
re-detected in the cleaned map, we consider them spurious. 

Furthermore, we compare the MHW source detections with the detections 
from the CLEAN algorithm (also shown in Table~\ref{tab:mhwclean}).  
The CLEAN algorithm detects twelve sources, of which eight are in 
good agreement with those detected by the MHW algorithm.  CLEAN yields 
larger flux estimates for sources 3 and 6.  The positions deviate 
by about 1-2 pixels from the MHW position, but in general there is good 
agreement between the two methods.  Source 7 is not detected by CLEAN.  
The four additional sources (13-16) are faint, $\sim$\,2\,mJy, and 
remain undetected by MHW.  The reason for this could be found in how the 
two methods depend on the zero-point of the map:  If large scale variations 
across the map elevate the zero-point of the map, this will also have an 
impact on the $S/N$ value so that these appear systematically off-set by 
a small value.  For the CLEAN routine, which selects the sources by 
$S/N>3$, also positions with artificially increased $S/N$ ratio will be 
selected.  The MHW algorithm, which searches for sources of a given scale 
placed on top of a background, such large scale variations will have no or 
only very small influence on the detection of sources, and thus MHW remains 
essentially insensitive to the zero-point of the map.  
For future large surveys, such as those planned to be carried out with 
SCUBA-2, large scale variations can pose a problem.  MHW may offer a 
good solution for source extraction from such maps.  

Finally, in order to check the MHW source detection, 
in order to check for possible sources hidden in the wings of bright 
sources, we subtracted the detected sources from the map, using a template 
for the central beam scaled to the fluxes of the individual sources. This 
residual map is not the residual map from the CLEAN routine, as it still 
contains the sources which were detected by CLEAN but not by MHW.  
Then, we performed source extraction with MHW on the residual image.  If 
there is a source closer than a beam to the other sources, which has 
remained undetected, then we expect to detect it in the residual map.  
Four such sources are detected.  
Three of these are directly associated with bright sources and are likely 
artifacts resulting from an imperfect match of the bright source and 
the scaled beam subtracted from the map.  
The fourth source may be real, but is outside of the 7.7\,arcmin$^2$ 
trimmed region near the SW edge and thus no included in the analysis. 
So in conclusion we believe that our MHW catalogue
is robust and does not miss any sources to the detection limit of the data.

\subsection{Testing MHW with simulations}
\label{sect2:simuls}

\begin{figure*}
\center{\includegraphics[width=14cm]{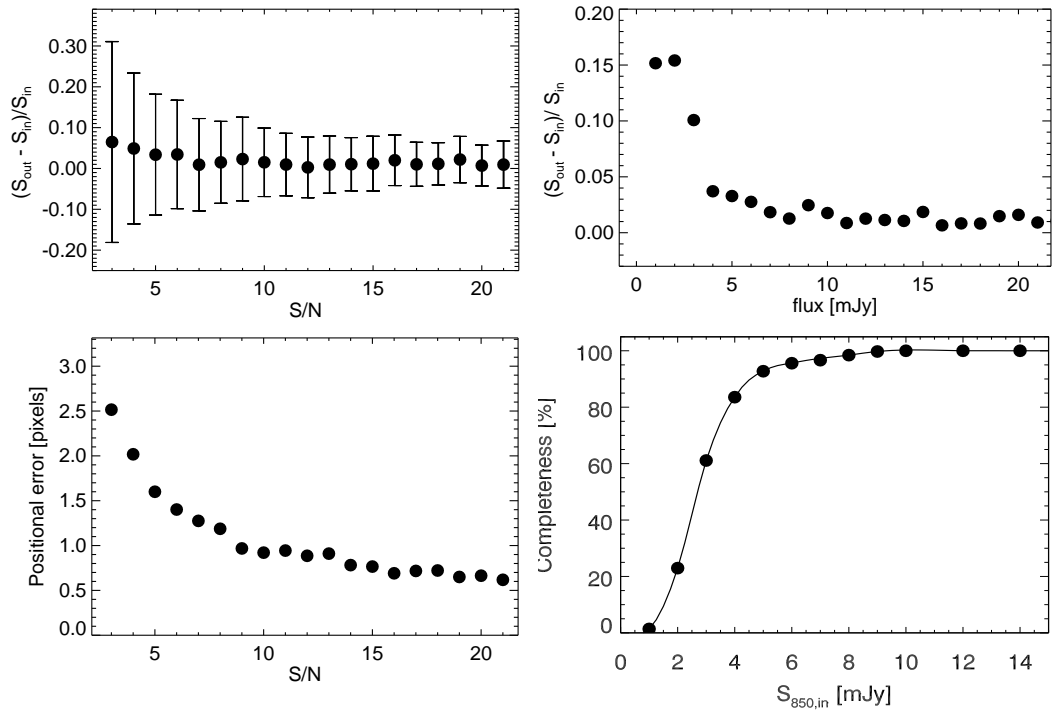}}
\caption[]{
{\it Upper panels:} \ 
The relative difference between the input and detected flux as function 
of $S/N$ ({\it left}) and flux ({\it right}) based on the simulations.  
The filled dots represent the average in each $S/N$ or flux bin, while 
the error bars represent the standard deviation in the same bin.  
{\it Lower left panel:} \ 
The standard deviation ($\bullet$) of 
the difference between input and detection position as a function of 
$S/N$ based on the simulations. 
{\it Lower right panel:} \ The 
completeness as function of input flux based on the 
simulations.   The 850\,$\mu$m map is 80\,\% complete at 3.8\,mJy 
and 50\,\% complete at 2.7\,mJy.  For comparison, 3\,mJy corresponds to 
$S/N\sim3$ in about two-thirds of the map. 
\label{fig2:res_sim}}
\end{figure*}
Using the empty Monte Carlo maps we investigated the detection accuracy of 
the MHW algorithm in a similar fashion as was performed for the scan-maps 
in \citet{barnard04}. 
One point source was added at a time to the empty maps using the central 
peak of the beam map shown in Figure \ref{fig2:beam} as a template source.  
This was repeated 400 times for each flux step. 
The position was randomly chosen, 
with a uniform distribution across ten maps, so that the whole map 
area was well sampled.  The input flux of the point source was then increased 
and the standard MHW detection algorithm was run on each map.
Additionally, we also perform two simulations for the CLEAN 
routine concerning sources close to the edge and overlapping sources.

\subsubsection{Position accuracy}
In Figure~\ref{fig2:res_sim} we plot the standard deviation of 
the difference between input and detection position.  The standard deviation 
decreases with increasing $S/N$ and has a value about 2.5\arcsec\ at 
$S/N\sim$\,3 decreasing towards 0.5\arcsec\ for large $S/N$.   
We adopt these as the uncertainty on the determined position introduced 
by the extraction algorithm.  
This added in quadrature with the pointing uncertainty of the JCMT 
($\sim$2\arcsec) gives the error on the determined position. 

\subsubsection{Flux accuracy} 
\label{subsect:fluxacc}
In Figure~\ref{fig2:res_sim} we plot the 
relative average flux difference, $(S_{out}-S_{in})/S_{in}$, and the 
standard deviation as function of $S/N$ and of flux.  
For $S/N > 7$ and also flux $>$\,7\,mJy the detected flux on average 
is overestimated by one to two percent.  Eddington bias is seen through 
the average overestimation of the sources with fainter fluxes and lower 
$S/N$.  
This is also seen in other SCUBA surveys (e.g., 
\citealt{scott_8mJy,webb_3h,borys03,barnard04}). 
We note that the calibration uncertainty is larger than the Eddington bias.  
The standard deviation on the average relative flux difference decreases 
with $S/N$, from about 25\% at $S/N=3$ to 5-7\% at $S/N>15$. 
As uncertainty on the determined flux we use the standard deviation 
of the relative average flux different, which is added in quadrature 
with the calibration error from the SCUBA reduction to give the total 
error on flux density of the individual sources. 

\subsubsection{Completeness}
The completeness as function of input flux was 
determined from the same set of simulations.  The fraction of sources 
detected at each flux level is calculated.  The result is plotted in 
Figure~\ref{fig2:res_sim}.   The observations are 80\,\% complete at  
3.8\,mJy and 50\,\% complete at 2.7\,mJy. 
 
\subsubsection{Sources near the edge}

For sources closer to the edge one of the two negative sidelopes will be 
outside of the map.  This can complicate the source detection in 
CLEAN, where a matching to the full beam pattern is done.  In simulating 
400 point sources, one at a time, each of 20 mJy, we found that 68 
sources, i.e.\ 17\%, were not detected, while similar sources away from 
the edge were all detected.  Additionally, the position and flux had 
larger uncertainties compared to those away from the edge. 

\subsubsection{Overlapping sources}

Given the large width of the 850\,$\mu$m beam, blending of sources 
is not an uncommon occurance.  For example in the A2218 
two of the nine sources are separated by approximately one beam width.  
We simulated source blending by adding two sources with a separation 
between 0 and 30 arcsec onto the MC maps.  
The flux ratio of the two sources was taken to be 1:1, 1:2 and 1:4, 
although we find that different flux ratios do not significantly 
alter the result in either case.  
MHW detects individual sources separated by more than one beam 
width, while CLEAN is able to detect sources with separations 
$>12\arcsec$.  However, the flux and positions accuracies using 
MHW are on average approximately 5\% better than those of CLEAN.


\section{Analysis of Abell 2218 sub-mm maps}     

\subsection{The sub-mm source catalogue}     

\begin{figure}
\center{\includegraphics[width=8cm]{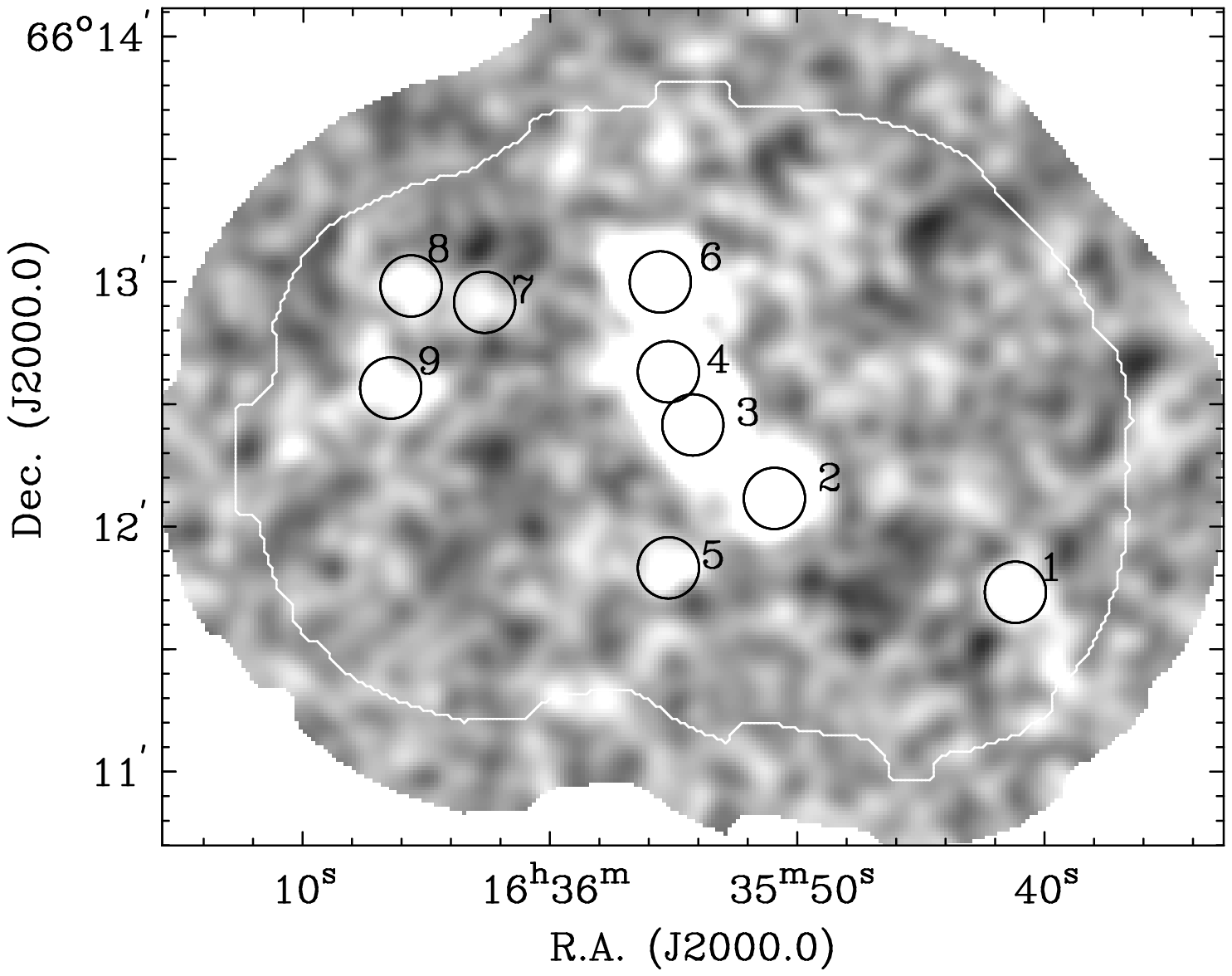}}
\center{\includegraphics[width=8cm]{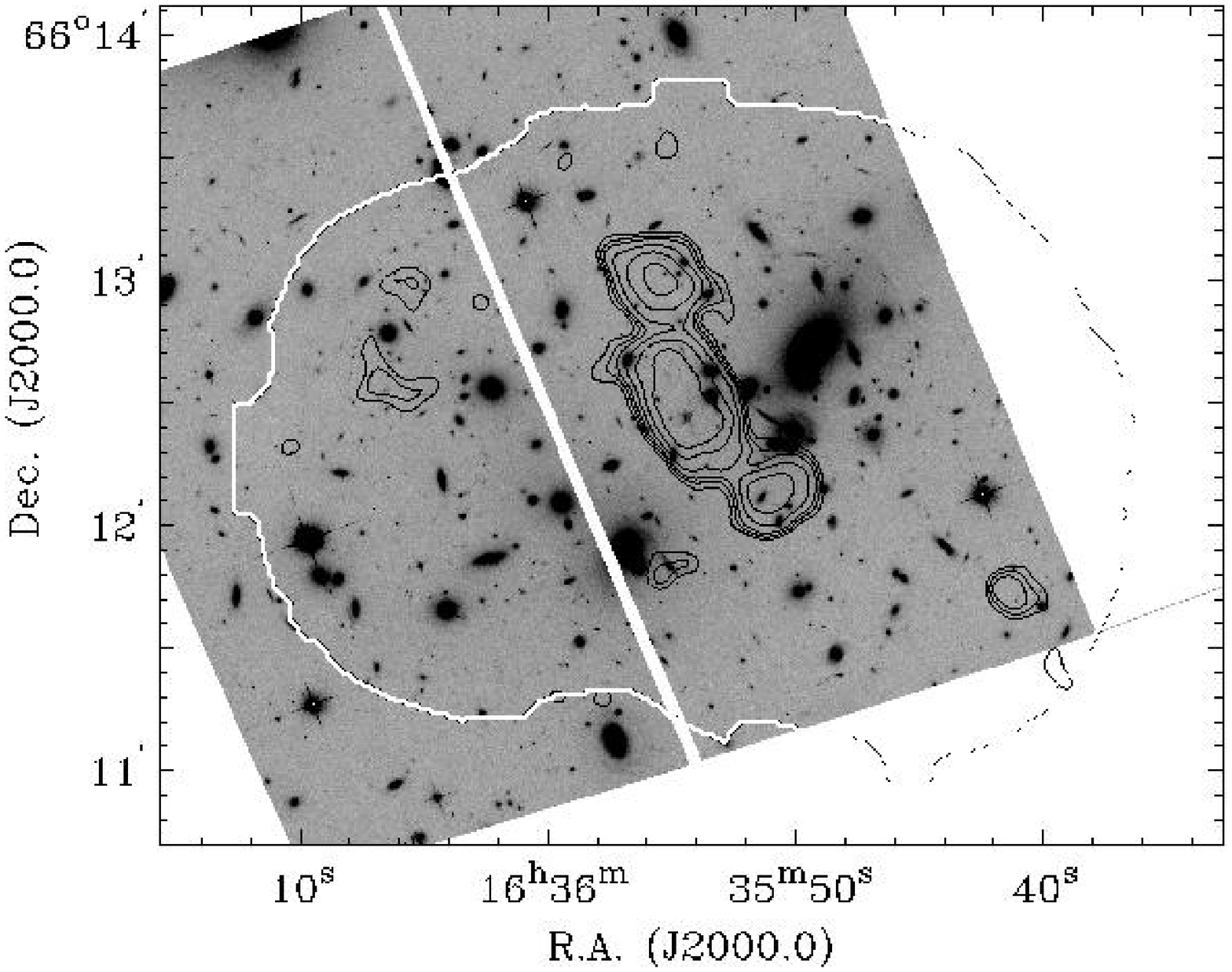}}
\caption[]{{\it Upper panel:} \ \ A2218 850\,$\mu$m $S/N$ map where the 
nine detected sources each are indicated with a circle with diameter 
of 15$''$.  The number assigned to each source corresponds to their 
order in Table \ref{tab2:results}. 
The white line indicates the area used for the 
analysis. 
{\it Lower panel:} \ \ The 850\,$\mu$m $S/N=3,4,5,7,10,15$ contours overlaid
on the HST ACS $z$-band image \citep{Kneib04b}.  
\label{fig2:a2218_850sn}}
\end{figure}

We now present the detailed analysis of the Abell 2218 SCUBA maps
using the MHW technique. We focus on a
total area of useful data of 7.7\,arcmin$^2$ (see figure \ref{fig2:850450mos}).  
Nine sources were detected by MHW using the detection parameters described
in Section 3.1. 
Six of these sources have detectable 450\,$\mu$m emission. 
The detected sources are listed in Table~\ref{tab2:results}.  
The sources are named using the prefix SMM followed by their J2000 coordinates. 
For the 850\,$\mu$m parameters we give the $S/N$ in real space, 
the combined flux and flux uncertainty and position uncertainty.  
For the 450\,$\mu$m, where no 
emission is detected, we give 3$\sigma$ upper limits.   
The uncertainty on the 450\,$\mu$m flux is 30\%, which is derived from the 
calibration of the absolute flux.  
In Figure~\ref{fig2:a2218_850sn} we show the 850\,$\mu$m $S/N$ map, and 
the $S/N$ contours overlaid on an optical image for orientation.
The differences seen in flux, in particular at 450\,$\mu$m, between these 
results and those of \citet{Kneib_a2218} can be assigned to uncertainties 
in the calibration.  More data has been included in the combined data set 
presented here than in the \citet{Kneib_a2218}.   

\begin{center}
\begin{table*}
\caption{Sources detected in the SCUBA maps of A2218. 
$f$ is flux; $S/N$ is the signal-to-noise of the source in real space as
detected using MHW; 
$\sigma$ gives the uncertainties on the flux and position at 850\,$\mu$m 
and 450\,$\mu$m respectively.  The flux uncertainty is that found for 
the appropriate $S/N$ bin added in quadrature with the calibration 
uncertainty, as described in section \ref{subsect:fluxacc}.  The 
positional uncertainty does not include confusion errors. 
\label{tab2:results} }
\vskip2mm
\begin{tabular}{lccccccccc}
\hline
\hline \\ [-2mm]
{\sc name} & & $\sigma_{pos}$ & $f_{850}\pm\sigma_{850}$ & $S/N$ & $f_{450}\pm\sigma_{450}$ & $S/N$ & $z_{spec}$ & $\mu$ & $f_{850}/\mu$ \\
           & &  $''$ & mJy & & mJy & mJy & & & mJy \\ [2mm]
\hline \\ [-3mm]
SMM\,J163541.2$+$661144         & (1) & 2.9 & 10.4$\pm$1.4 & 7.5  & $<59.4$       & ...  & 3.188 &  1.7 & 6.0  \\
SMM\,J163550.9$+$661207$^\star$ & (2) & 2.7 &  8.7$\pm$1.1 & 11.5 & 39.2$\pm$11.8 & 7.2  & 2.516 & 45$^\star$ & 0.8  \\
SMM\,J163554.2$+$661225$^\star$ & (3) & 2.6 & 16.1$\pm$1.6 & 21.7 & 89.7$\pm$26.9 & 19.0 & 2.516 & 45$^\star$ & 0.8  \\
SMM\,J163555.2$+$661238$^\star$ & (4) & 2.6 & 12.8$\pm$1.5 & 16.9 & 60.1$\pm$18.3 & 12.1 & 2.516 & 45$^\star$ & 0.8  \\
SMM\,J163555.2$+$661150         & (5) & 3.5 &  3.1$\pm$0.7 & 3.8  & 29.1$\pm$8.7  & 6.1  & 1.034 &  7.1 & 0.4  \\
SMM\,J163555.5$+$661300         & (6) & 2.6 & 11.3$\pm$1.3 & 15.8 & 18.0$\pm$5.4  & 3.0  & 4.048 &  4.2 & 2.7  \\
SMM\,J163602.6$+$661255         & (7) & 3.5 &  2.8$\pm$0.6 & 3.5  & $<21.6$       & ...  & ...   &  1.8 & 1.6  \\
SMM\,J163605.6$+$661259         & (8) & 3.3 &  5.2$\pm$0.9 & 4.9  & $<25.5$       & ...  & ...   &  1.8 & 3.3  \\
SMM\,J163606.5$+$661234         & (9) & 3.3 &  4.8$\pm$0.8 & 4.6  & 30.1$\pm$9.3  & 3.1  & ...   &  1.6 & 2.9  \\ [1mm]
\hline
\end{tabular}
\flushleft{$^\star$ SMM\,J163550.9$+$661207, SMM\,J163554.2$+$661225 and SMM\,J163555.2$+$661238  
have been identified as a multiply-imaged galaxy, which is denoted 
SMM\,J16359+6612 \citep{Kneib_a2218}.}
\end{table*}
\end{center}

\subsection{Gravitational Lensing Magnification}
\begin{figure}
\center{\includegraphics[width=7.1cm]{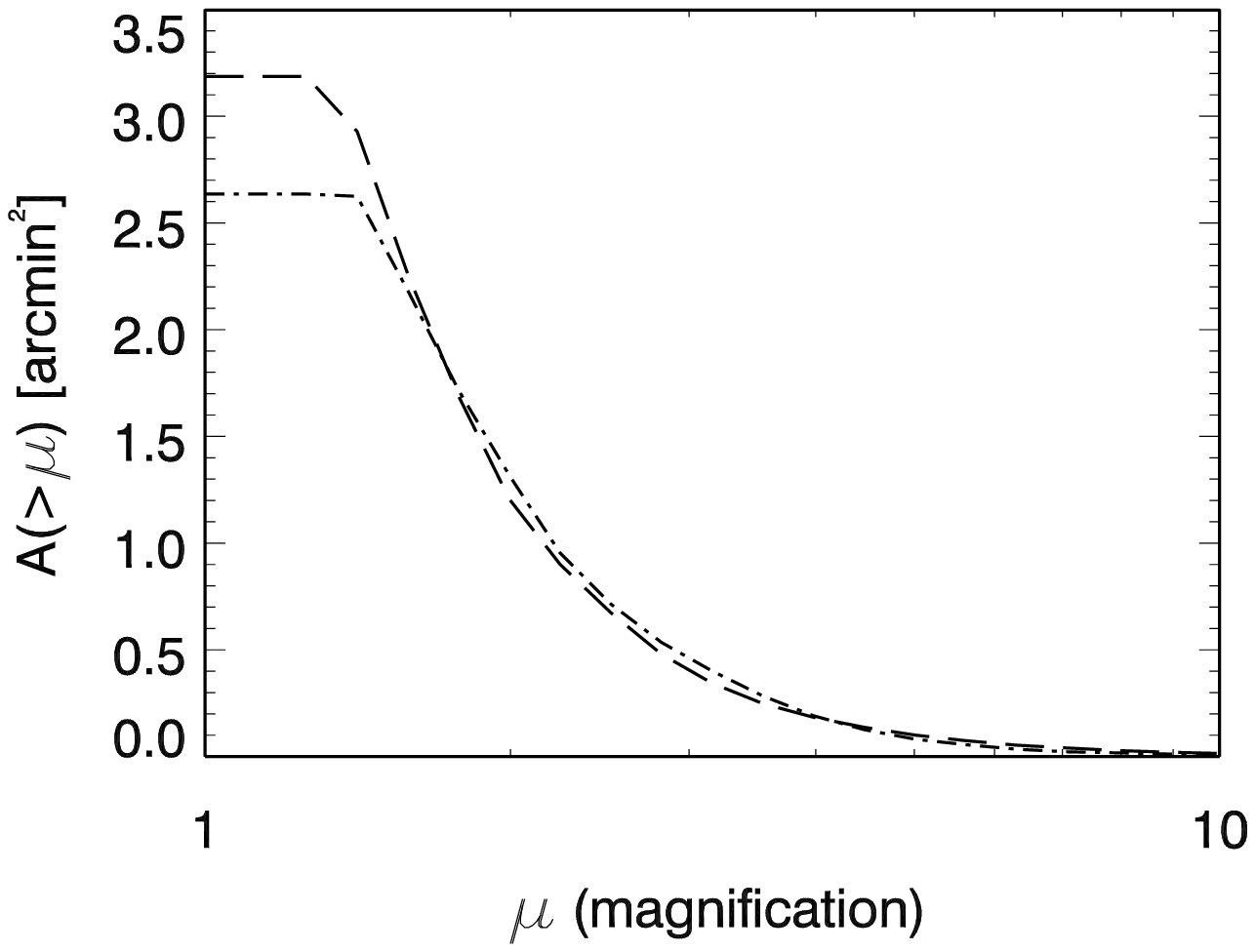}} \center{\includegraphics[width=7.1cm]{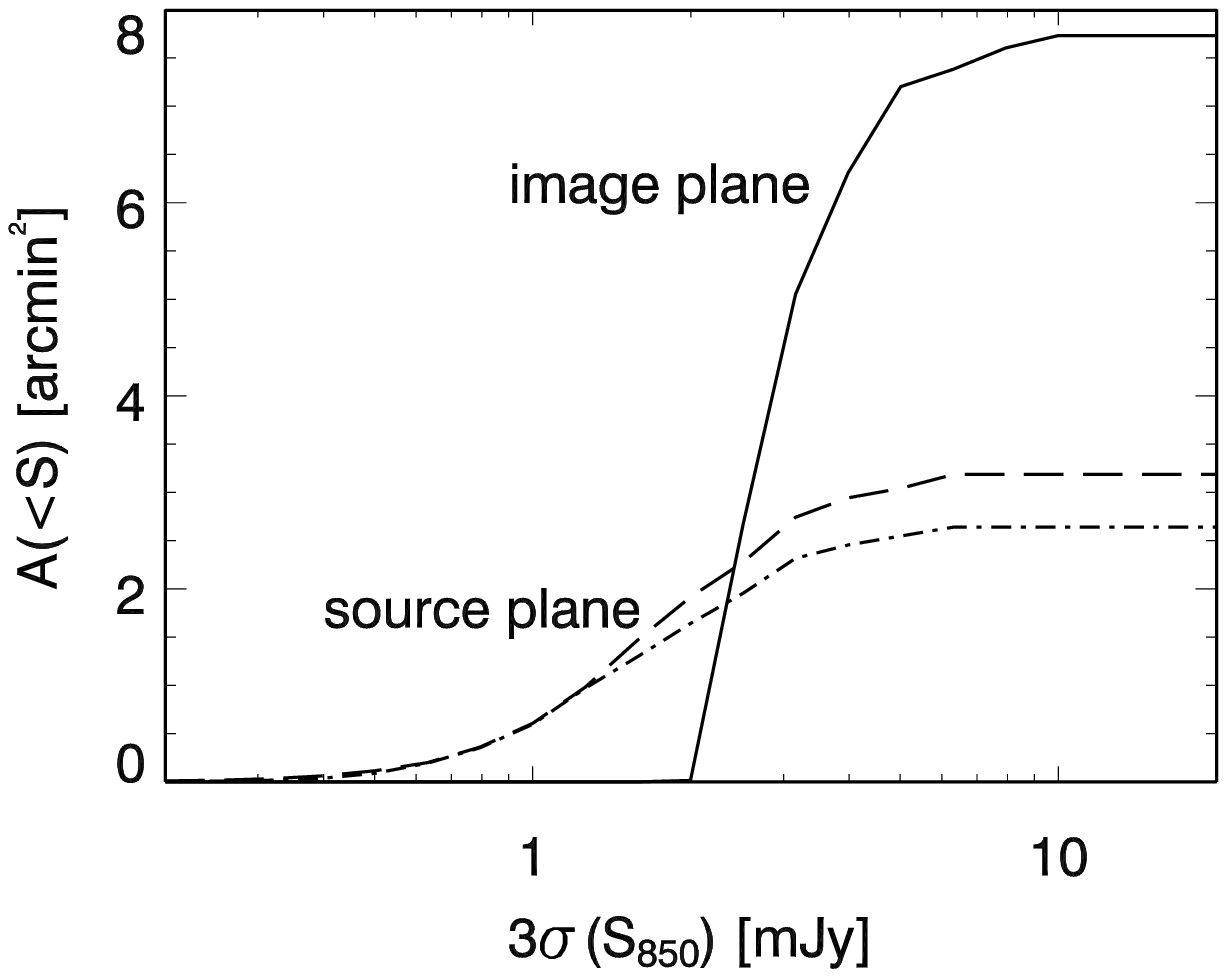}} \center{\includegraphics[width=7.1cm]{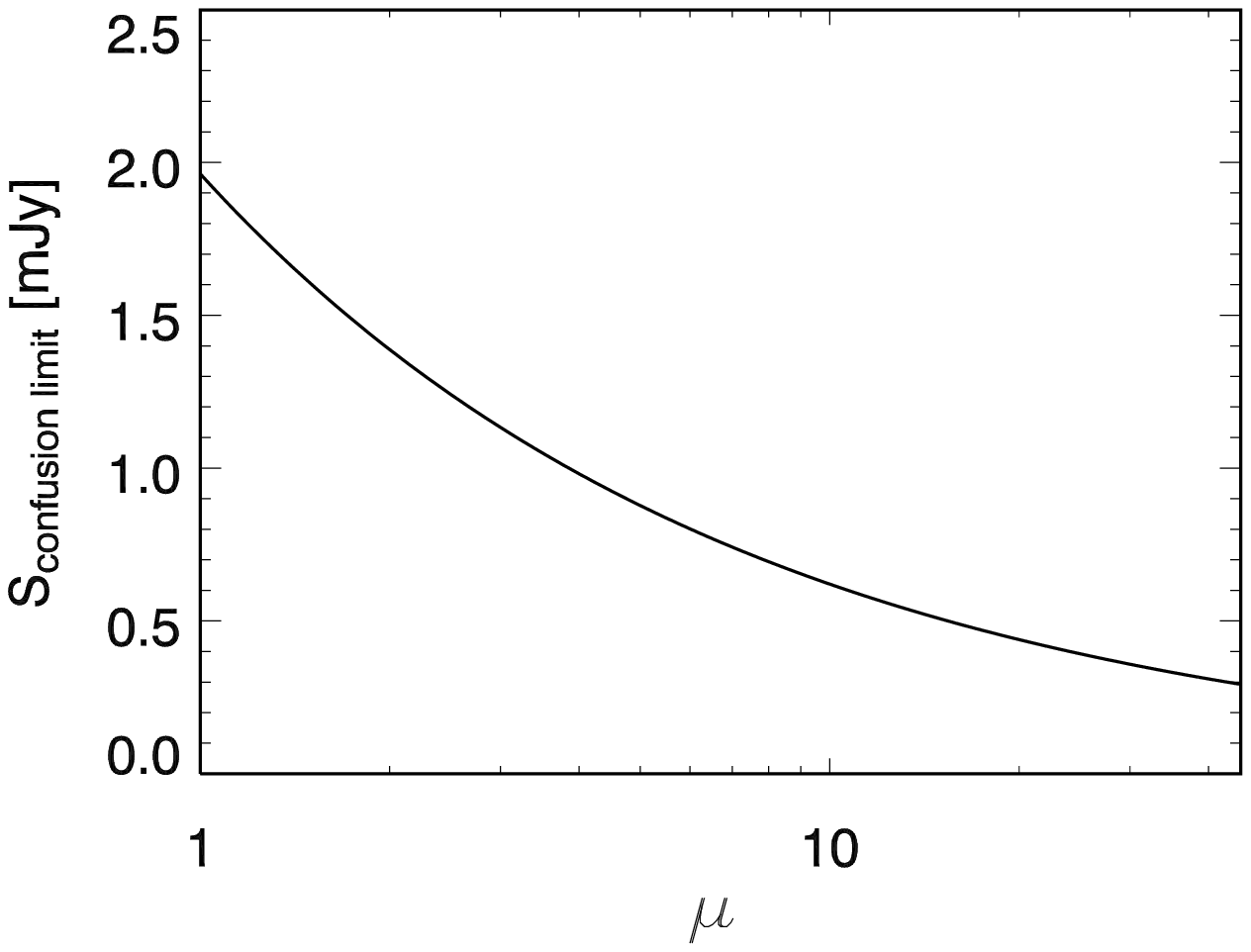}}
\caption[]{
{\rm Upper panel:} \ \ The area of the source plane as function of 
magnification for redshifts $z= 1$ (dashed) and $z = 4$ (dash-dot).  
The difference between the different redshifts is relatively small.  
{\rm Middle panel:} \ \ The area of the source plane as function of 
3$\sigma$ sensitivity
for redshifts $z= 1$ (dashed) and $z = 4$ (dash-dot).  
The source plane at $z=2.5$ 
is about 2.8 times smaller than the observed area of the image plane.  
{\rm Lower panel:} \ \ The confusion flux limit as function of 
magnification, as described in subsection \ref{subsect2:conf}. 
\label{fig2:lensarea}}
\end{figure}

We exploit the detailed mass model of A2218 \citep{kneib96}
updated to include the triple sub-mm image \citep{Kneib_a2218}
and the high redshift multiple images
at $z= $ 5.56 (Ellis et al., 2001) and $z\sim 7$ (Kneib et al 2004b).
In Figure~\ref{fig2:lensarea} we show the area as function of magnification 
for source planes at the redshifts $z= $ 1 and 4.
In 1.3\,arcmin$^2$ the flux magnification factors are $>$\,2. 
Furthermore, in Figure~\ref{fig2:lensarea} we show the area as function of 
sensitivity both in the image plane and in the source plane, 
again for source planes at different redshifts, $z= $ 1 and 4. 
For high redshift, $>1$, the lensing magnification is only weakly 
depending on redshift. 
For the sources without known redshift, assuming a redshift $>1$ will 
not have a strong impact on the derived properties. 
For $z>2$, the effective area surveyed (source plane) is about
2.7\,arcmin$^2$ which corresponds to an average magnification factor of 2.8.

The redshift is known for six sources in the field:  three of these are 
a multiply-imaged object at redshift $z= $ 2.516 (Kneib et al., 2004a), 
while the other three are single-imaged sources at 
redshift $z= $ 1.034 \citep{pello92}, and $z=3.188$ and 4.048 
(Paper II).  
For the sources with unknown redshift we assume $z= $ 2.5,  based on 
the median redshift for bright SCUBA sources determined by 
\citet{chapman03,chapman05}. 
The magnification factors of the individual singly lensed
sources range between 1.6 and 7.1. For the multiply-imaged 
source at $z= $ 2.516 the magnification factors for individual 
images are 9, 14 and 22, i.e.\ a total magnification of 45
for the three images combined (Kneib et al., 2004).
In Table~\ref{tab2:results} we list the magnification factors, $\mu$, 
and the lensing corrected fluxes, $f_{850}/\mu$ for the individual galaxies.  

Note that three (five, including all images of SMM\,J16359+6612) 
of the nine detected sources have unlensed fluxes 
fainter than or close to the blank field confusion limit.  
These sources would most 
likely not be detected in blank field surveys and thus not have been 
accessible to us without the use of gravitational lensing or a larger 
telescope with sensitive instruments.

\subsection{Confusion}
\label{subsect2:conf}

Confusion noise is caused by unresolved faint sources in the field. 
In the background of deep SCUBA maps the instrumental noise and the
confusion noise from a fainter submm population are of approximately
equal magnitude.
Understanding the confusion in a map is essential to be able to estimate 
the relative importance of uncertainties due to blending of sources.  

For the cluster fields the confusion limit is affected by the
gravitational lensing.
The gravitational lensing magnifies the region seen behind the
cluster, hence the source plane is smaller than the image plane.
The number of beams is conserved between the image plane and the
source plane, i.e.\ the size of the beam scales with the magnification.
This is why it is at all possible to observe the fainter sources, which
have a higher surface density than the brighter sources.

The number counts in the lensed case can be written as
$N_{lens}= (N_0 / \mu) (S / \mu)^{-\alpha}$
$= N_{blank} \mu^{\alpha-1}$,
where $\mu$ is the gravitational lensing magnification, 
assuming a blank-field number count power law where $N(>S) = N_0 S^{-\alpha}$ 
and $S$ is the observed flux.
The average magnification for a field can be found as the ratio
of the area in the image plane and the area in the source plane.
We use the rule of thumb, that the confusion limit in imaging 
is defined as one source per 30 beams (e.g.\ \citealt{hogg}). 
The confusion limit in the lensed case can thus be written as
$S_{conf} = (30\Omega N_0 \mu^{1-\alpha})^{1/\alpha}$ 
(illustrated in Figure~\ref{fig2:lensarea}), where $\Omega$ is the 
solid angle of the beam. 
As the lensing magnification varies across the field, this means that the 
actual confusion limit is deeper in highly magnified regions close to 
the caustics than the estimated average lensed confusion limit. 

In case of the A2218 850\,$\mu$m map, the confusion limit in the
source plane based on this simple estimate is $\sim$\,1.1\,mJy on 
average across the whole field. 
In the simplified estimate of the confusion limit presented here
we assumed that the number counts are described by a
single power law.  There are good indications that the number counts
are described by a double power law or another function with a (gradual)
turn-over (\citealt{ScottWhite99,scott_8mJy,borys03,KnudsenPHD}; Knudsen et al., {\it in prep.}).   
In the Leiden-SCUBA Lens Survey (\citealt{KnudsenPHD}), the 850 number counts 
are probed over almost two decades of flux (0.1-20\,mJy) and improving 
the statistics in particular on the faint end.
Including double power law number counts in such a calculation will 
work in a favourable
direction and the confusion limit in the source plane will be lower.

\subsection{850\,$\mu$m source counts in A2218}

\begin{figure*}
\includegraphics[width=8.5cm]{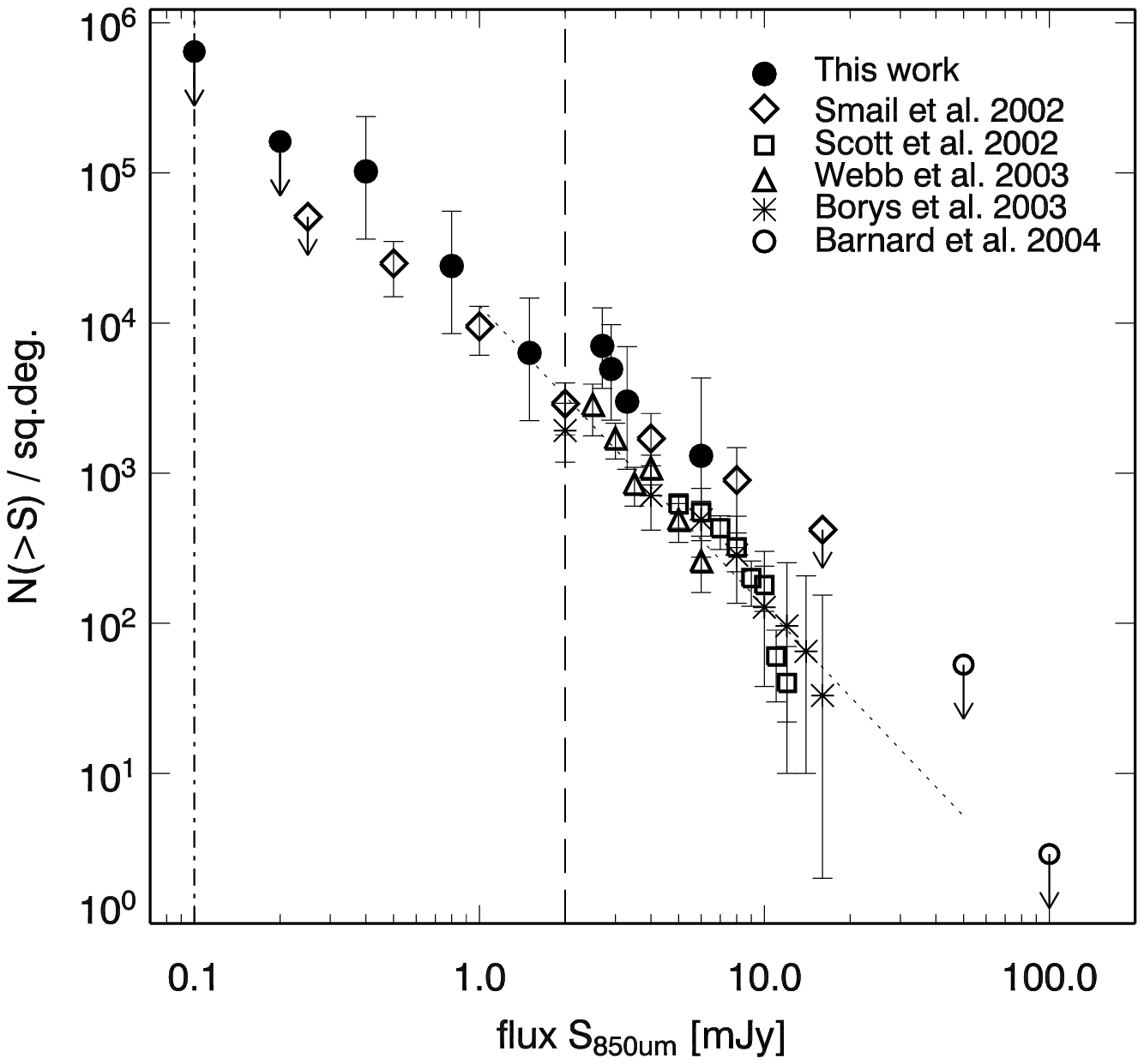}
\includegraphics[width=8.5cm]{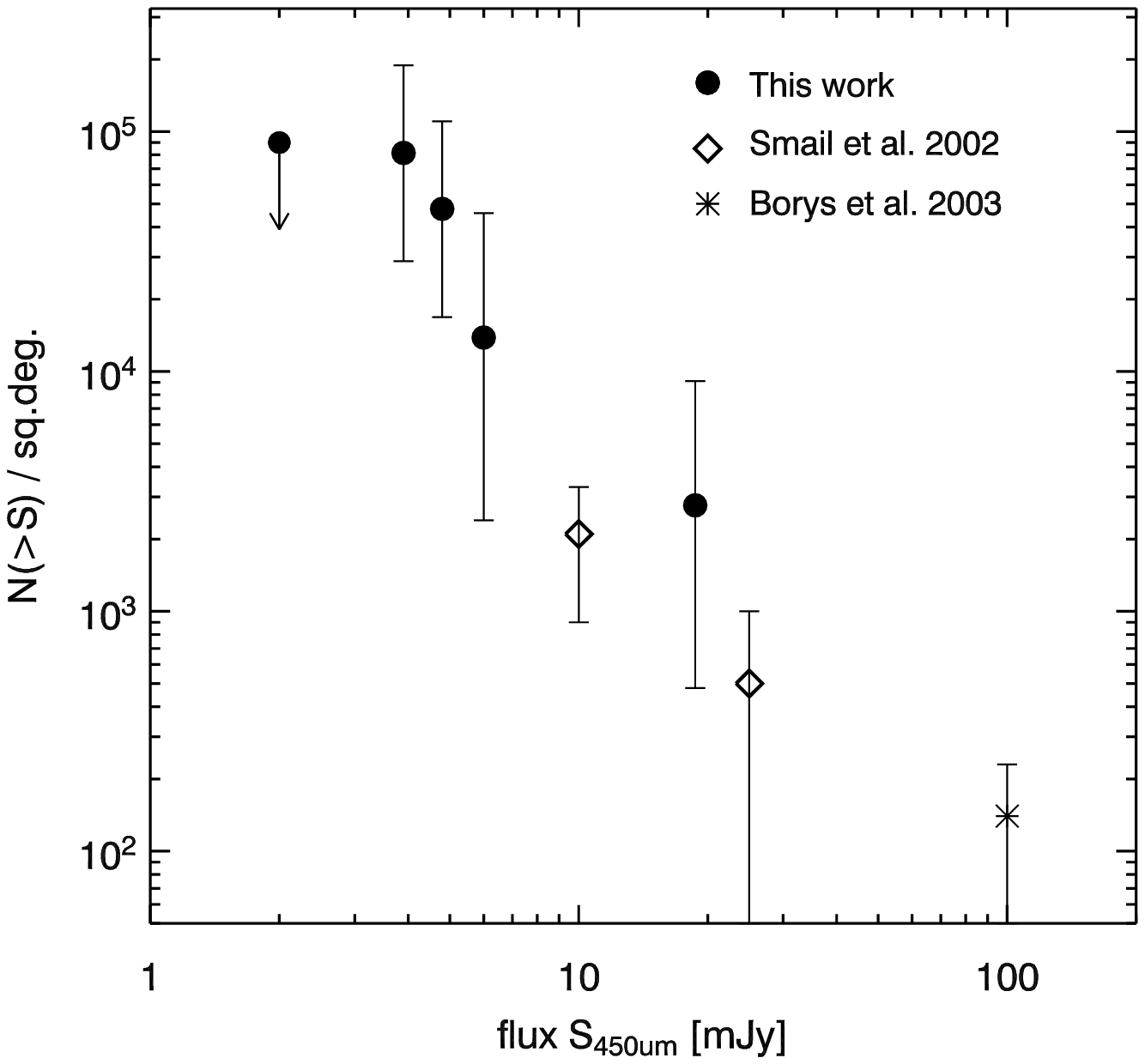}
\caption[]{The 850\,$\mu$m (left) and 450,$\mu$m (right)
source counts in A2218 (filled circles). 
The upper limits for the very bright number counts from \citealt{barnard04} 
are also shown (open circles). 
Furthermore, the number counts from four other surveys are included: 
the Lens SCUBA survey (diamonds; \citealt{sibk02}), the 8mJy-survey 
(squares; \citealt{scott_8mJy}), the Canada-UK Deep SCUBA Survey 
(triangles; \citealt{webb_3h}) and the HDF Super-map (stars; 
\citealt{borys03}). 
The dotted line indicates the power-law counts used for calculating the 
confusion limit in subsection \ref{subsect2:conf}.  The dashed line 
shows the blank field confusion limit. 
The vertical dash-dotted line indicate the 5$\sigma$ detection limit 
for a two hours observations with ALMA.  
\label{fig:a2218counts}}
\end{figure*}

In Fig.~\ref{fig:a2218counts} we plot the number counts for sources 
detected in A2218.   The number counts are calculated for the 
flux levels corresponding to the flux of each of the sources.  
Because of the large variation in the sensitivity across the map, 
it is not possible to directly determine the number counts as 
cumulative number counts.  The complications arising from this has 
been discussed in previously (e.g., \citealt{borys03,webb_3h}).  
We here apply a simple method for estimating the number counts:  
At each flux level, $S$, only the area where $S > 3\sigma$ are 
considered.  The number of sources within that area are divived by 
the area. 
We use the sensitivity map for the source plane at redshift $z=2.5$ 
and note that using that of sources planes 
at other redshifts does not make a noticable difference.  
By using the source plane sensitivity map, we also take into account 
the effect that lensing has on the area.  
The error bars are Poisson statistics \citep{gehrels}.   
The number counts are in 
reasonable agreement with the results from other surveys. 
Differences between surveys can be assigned to cosmic variance and 
small number statistics.   
The 850\,$\mu$m
number counts at the level of $\sim 1$\,mJy is $\sim 2$\,arcmin$^{-2}$. 
Additionally, we determine an upper limit for the number counts at 
$S_{850} = 0.2$\,mJy and 0.1\,mJy of 44\,arcmin$^{-2}$ and 175\,arcmin$^{-2}$ 
respectively.   
The 450\,$\mu$m number counts probe deeper than previously 
published \cite{sibk02,borys03}.  They appear to be slightly higher, which can 
be assigned to sample variance, but agree within the error bars. 

The fraction of the submm extragalactic background light (EBL) resolved 
in the SCUBA map can be constrained through the sum of the fluxes. 
The total flux relative to the area is (35$\pm$1.6)\,Jy\,deg$^{-2}$.  This 
is $\sim$\,80\% of the EBL as determined by \citet{fixsen98}, and 
comparable to the EBL value determine by \citet{puget96}.  
This almost double the total flux value of 17.3 Jy\,deg$^{-2}$ 
found by \citet{cowie}.   Due to cosmic variance and the clustering 
of SCUBA sources, this value is expected to vary between fields. 
However, the fact that we resolve such a significant fraction of the 
EBL, is a good demonstration of the power of gravitational lensing 
in combination with the SCUBA observations, as it allows us to 
probe the faint submm galaxy population, which is the dominant 
contributor to the submm EBL.  


\section{Conclusions and outloook}     

We have performed a thorough analysis of the Mexican Hat Wavelet  
algorithm (MHW) as a source extraction tool applied to SCUBA jiggle-maps. 
The analysis was done using the deep SCUBA maps of the galaxy cluster 
A2218,  through source extraction with MHW on the real 
data and through extensive simulations.  We found that MHW is a stable 
method for source extraction at low signal-to-noise ($S/N > $ 3). 
We conclude that MHW is an algorithm suitable for source extraction 
from SCUBA jiggle maps and is a powerful tool for studying the 
faint submillimetre sources and thereby the faint end of the submillimetre number counts. 
MHW has potential as source extraction algorithm for data taken 
with the future SCUBA-2 instrument.  SCUBA-2 will make ``total power''
maps, which should be free of chopping artifacts, hence MHW will have 
an immediate advantage over the other techniques currently in use. 

The SCUBA map we have obtained for A2218 is one of the deepest submillimetre maps ever taken.
Covering a large area of the cluster, we have been able to survey the 
region where strongly lensed background sources are present.  
In the analysis of A2218, nine sources were detected in the 850\,$\mu$m 
SCUBA map, the largest number detected in a single ultra-deep SCUBA map.  
Six of these sources were also detected at 450\,$\mu$m.  
Correcting for the gravitational lensing by the galaxy cluster, three 
sources have intrinsic 850\,$\mu$m fluxes below the blank field 
confusion limit, and three with fluxes comparable to the blank field 
confusion limit. 
It is the 
presence of strong gravitational lensing, which pushes the confusion 
limit to fainter fluxes levels, which has made it possible to detect 
the faint submillimetre source with \mbox{$< $ 2\,mJy}. 
For this field we determine the 850\,$\mu$m
number counts to $S_{850} = 0.4$, and 
place  upper limits at $S_{850} = 0.1$ and 0.2\,mJy.  Additionally, 
we also estimate the 450\,$\mu$m number counts down to $S_{450} = 4$\,mJy.
The identification of the individual sources will follow in Paper II. 

ALMA will in the future allow for yet deeper observations and thus 
a continuation of the number counts to even fainter fluxes.  If ALMA 
will detect a number density of $\sim 50$ galaxies/arcmin$^2$, this 
means we are reaching the number density of galaxies detected in the 
optical/near-infrared and thus will allow for a better understanding 
of the connection between dust and stars in the Universe as 
function of time/redshift. 

\section*{acknowledgements}

  We thank Tracy Webb, George Miley and Douglas Pierce-Price for useful discussions. 
  We also thank our anonymous referee for constructive suggestions, which 
  helped improve the manuscript. 
  The JCMT is operated by the Joint Astronomy Centre on behalf of the
  United Kingdom Particle Physics and Astronomy Research Council (PPARC),
  the Netherlands Organization for Scientific Research and the National
  Research Council of Canada.
  KKK acknowledges support from the Netherlands Organization for Scientific
  Research (NWO) and  the Leids Kerkhoven-Bosscha Fonds for travel support.
  JPK acknowledges support from Caltech and CNRS. 
  AWB acknowledges support from NSF grant AST-0205937, the Research 
  Corporation and the Alfred P.\ Sloan Foundation.
  IRS acknowledges support from the Royal Society.

\setlength{\bibhang}{2.0em}

\label{lastpage}

\end{document}